\documentclass[epj]{svjour}

\usepackage{graphicx}
\usepackage{bm}

\usepackage{epsfig}
\usepackage{graphics}
\usepackage{graphicx}
\usepackage{amssymb}

\usepackage{color}
\newenvironment{ckomm}{}{ }
\newcommand{\QQ}{\begin{ckomm}\color{red} \bf}
\newcommand{\QQEND}{\end{ckomm}}

\newcommand{\kb}[0]{k_{\mathrm{B}}}
\newcommand{\el}[0]{(E,\,L)}
\newcommand{\dd}[0]{\mathrm{d}}
\newcommand{\ee}[0]{\mathrm{e}}
\newcommand{\rr}[0]{\mathbf{r}}
\newcommand{\pp}[0]{\mathbf{p}}

\newcommand{\coll}[0]{\Gamma_\mathrm{coll}}
\newcommand{\moy}[1]{\left \langle #1 \right \rangle}
\newcommand{\ket}[1]{\left | #1 \right \rangle}

\newcommand{\eperp}[0]{\langle E_\bot\rangle}

\begin{document}
\title{Discrete-step evaporation of an atomic beam}
\titlerunning{Discrete-step evaporation of an atomic beam}

\author{T. Lahaye and D. Gu\'ery-Odelin}
\institute{Laboratoire Kastler Brossel\thanks{Unit\'e de Recherche
de l'Ecole Normale Sup\'erieure et de l'Universit\'e Pierre et
Marie Curie, associ\'ee au CNRS.
}, D\'epartement de Physique de l'Ecole Normale Sup\'{e}rieure, \\
24 rue Lhomond, 75005 Paris, France}

\date{\today}

\abstract{We present a theoretical analysis of the evaporative
cooling of a magnetically guided atomic beam by means of discrete
radio-frequency antennas. First we derive the changes in flux and
temperature, as well as in collision rate and phase-space density,
for a single evaporation step. Next we show how the occurrence of
collisions during the propagation between two successive antennas
can be probed. Finally, we discuss the optimization of the
evaporation ramp with several antennas to reach quantum
degeneracy. We estimate the number of antennas required to
increase the phase-space density by several orders of magnitude.
We find that at least 30 antennas are needed to gain a factor
$10^8$ in phase-space density.}


\maketitle

\section{Introduction}
\label{sec:introduction}

Evaporative cooling \cite{KetterleVanDruten} is so far the only
technique allowing the achievement of quantum degeneracy in dilute
gases. Several models of evaporative cooling of a trapped cloud of
atoms have been studied. They are based either on approximating
the cooling process as a series of truncations of the distribution
function followed by rethermalization \cite{ketterleEvap}, or on
an approximate analytical solution of the Boltzmann equation
\cite{walravevap}.

With the achievement of Bose-Einstein condensation (BEC), the
possibility of realizing sources of coherent matter waves has
arisen. Those ``atom lasers" \cite{atomlaser} open the way to
fascinating applications in atom optics and interferometry. To
date, all atom lasers have been achieved in a pulsed mode:
coherent streams of atoms were extracted from a Bose-Einstein
condensate until it was completely depleted. As a first step
towards the realization of a cw atom laser, a continuous source of
Bose-Einstein condensed atoms was created by periodically
replenishing a condensate held in an optical dipole trap with new
condensates \cite{ScienceK02}. This approach, in combination with
an appropriate outcoupler, would lead to a cw atom laser.

Alternatively, the authors of \cite{Mandonnet00} study the
evaporation of an atomic beam propagating in a magnetic guide.
They use an approximate solution of the Boltzmann equation based
on a truncated gaussian ansatz \cite{walravevap}. This approach
permits the establishment of a set of hydrodynamic-like equations
which is solved numerically. This treatment however can be made
rigorous only for a one-dimensional evaporation. It would describe
correctly evaporation on a dielectric surface on which atoms can
be adsorbed \cite{cornell}.

From an experimental point of view, the propagation of a single
packet of atoms has been realized in both macroscopic and
microscopic atom guides
\cite{Schmiedmayer95,Denschlag99,Goepfert99,Key00,Dekker00,Teo01,Sauer01,Hinds99}.
More recently, a continuous magnetically guided beam was achieved,
by directly injecting a beam of cold atoms generated by a moving
molasses magneto-optical trap \cite{cren} as well as by feeding a
magnetic guide periodically at a high repetition rate
\cite{prlrb2}. A first step of evaporation on this continuous
guided beam has even been carried out by means of a single
radio-frequency antenna. A natural way to attain degeneracy
consists in using several radio-frequency (RF) antennas with
decreasing frequencies. It would be the analog in space of the
time-dependent RF ramp used in standard BEC experiments. The major
difference lies in the fact that the evaporative cooling is
ensured by successive cycles of evaporation followed by
rethermalization.

In this work, we present the corresponding discrete model of
evaporation by successive antennas. The two-dimensional character
of the confinement allows us to derive analytical expressions that
can be directly compared to experiments. After describing the
magnetic guide in which atoms propagate, we derive how a single
evaporation step affects the parameters of the beam (flux,
temperature) and we deduce how the elastic collision rate within
the beam, as well as the on-axis phase-space density, evolve. We
then show how one can characterize the occurrence of
rethermalization between two successive antennas as was
demonstrated experimentally in \cite{prlrb2}. We finally address
the issue of the optimization of the evaporation ramp with many
antennas.

\section{Magnetic guide}
Quadrupole guides can be produced by means of four wires or tubes
equally spaced on a cylinder of radius $a$ and carrying currents
$+I$ and $-I$ alternatively. The cylindrical symmetry axis is
chosen to be the $z$-axis in the following. The resulting magnetic
field is well approximated by a linear form: $B(r)=br$ where
$r=(x^2+y^2)^{1/2}$ is the distance from the $z$-axis and
$b=2\mu_0 I /(\pi a^2)$ is the transverse gradient. In this
quadrupolar configuration, atoms with low angular momentum are not
stable against spin flips. To counteract this loss mechanism, one
superimposes a bias field $B_0$ along the axis of the guide
\cite{quadnot}. As a consequence, the potential experienced by the
low-field seeking atoms is:
\begin{equation}
U_g(x,y,z)=\mu[B_0^2+b^2r^2]^{1/2}, \label{potential}
\end{equation}
where $\mu$ is the magnetic moment of an individual atom.

In the following, we consider a beam of atoms of mass $m$,
transversally confined by the potential (\ref{potential}), with a
mean velocity $\bar{v}$, a temperature $T$, and a flux $\Phi$.

At equilibrium the thermal average of  $U_g$ is:
\begin{equation}
\langle U_g\rangle_T=\frac{\int_0^\infty U_ge^{-\beta U_g}r\,\dd
r}{\int_0^\infty e^{-\beta U_g}r\,\dd
r}=k_BT\bigg(2+\frac{\alpha^2}{1+\alpha} \bigg), \label{eq:poten}
\end{equation}
where $\beta=1/\kb T$ and $\alpha=\mu B_0/k_BT$. For rubidium 87
polarized in the low-field seeking state of the lowest hyperfine
level $\ket{F=-m_F=1}$, $\mu=\mu_B/2$ where $\mu_B$ is the Bohr
magneton, and $\alpha=1$ with $B_0=1$~Gauss for a temperature of
$T=34\;\mu$K.

In the context of evaporative cooling, two quantities are of
interest: The on-axis phase-space density ${\mathcal D}$ and the
elastic collision rate $\coll$. For the potential
(\ref{potential}) we readily obtain
\begin{equation}
{\mathcal D}= \frac{1}{2\pi}\,\frac{1}{1+\alpha}\,\frac{\Phi}{\bar
v}\,\left(\frac{\mu b}{\kb T}\right)^2\,\frac{h^3}{(2\pi m \kb
T)^{3/2}}\,. \label{eq:psd}
\end{equation}
Assuming that the elastic collision cross-section $\sigma$ is
constant the collision rate reads:
\begin{equation}
\coll=\frac{\sigma}{2\pi^{3/2}}\,\frac{1+2\alpha}{(1+\alpha)^2}\,\frac{\Phi}{\bar
v}\,\left(\frac{\mu b}{\kb T}\right)^2\sqrt{\frac{\kb T}{m}}\,.
\label{eq:cr}
\end{equation}

Two limits are of interest depending on the value of $\alpha$. In
the low temperature regime ($\alpha\gg 1$), one can perform a
Taylor expansion of (\ref{potential}) resulting in the harmonic
form with an energy offset $\mu B_0$:
\begin{equation}
U_g(x,y,z)\simeq \mu B_0+ \frac{1}{2}\,m\omega^2r^2,
\label{potharm}
\end{equation}
and angular frequency $\omega=[\mu b^2/(mB_0)]^{1/2}$. The thermal
potential energy (\ref{eq:poten}) reduces to $\mu B_0+\kb T$, and
${\mathcal D}$ (resp. $\coll$) scales with $\Phi$ and $T$ as $\Phi
T^{-5/2}$ (resp. $\Phi T^{-1/2}$). In the opposite limit, where
$\alpha\ll 1$ one deals with a linear potential:
$U_g(x,y,z)\simeq\mu br$. The thermal potential energy is
$\moy{U_g}_T=2k_BT$, and the scalings for ${\mathcal D}$ and
$\coll$ are $\Phi T^{-7/2}$ and $\Phi T^{-3/2}$.

\section{Evaporation with one antenna}
\label{sec:1ant}

In order to perform evaporation on the beam, one uses an RF
antenna to produce an oscillating magnetic field at a frequency
$\nu_{\rm RF}$. Atoms whose trajectories cross the cylinder of
axis $z$ and radius $R$, defined by $h\nu_{\rm RF}=U_g(R)$, will
be resonant with the RF field and undergo a spin flip transition
to an untrapped state. They are therefore removed from the atomic
beam, whose flux decreases to the new value $\Phi'$.

We define the \emph{evaporation parameter} $\eta$ as $\eta =
h(\nu_{\rm RF}-\nu_{\rm 0})/\kb T$, where $\nu_{\rm 0}=U_g(0)/h$
is the frequency corresponding to the bottom of the trap. In the
following we first calculate, as a function of $\eta$, the
fraction $\phi(\eta)=\Phi'/\Phi$ of remaining atoms after such an
\emph{evaporation cycle}. The distribution function of the atoms
just after evaporation is out of equilibrium. Assuming subsequent
rethermalization of the beam, we then calculate its new
temperature $T'$. In this paper we assume that the collision rate
$\coll$ is much smaller than the typical period of oscillation
(radial collisionless regime). In other words, we assume that
particles do not undergo collisions over the range of efficiency
of a radio frequency antenna. Therefore, determining if a particle
will be evaporated or not depends only on the characteristics of
its trajectory.

\subsection{Fraction $\phi(\eta)=\Phi'/\Phi$ of remaining atoms}

To calculate $\phi(\eta)$, we need to know the fraction of atoms
in the beam whose trajectories in the $(xy)$ plane cross the
circle of radius $R$.

\subsubsection{Effective potential}

\begin{figure}
\begin{center}
\includegraphics{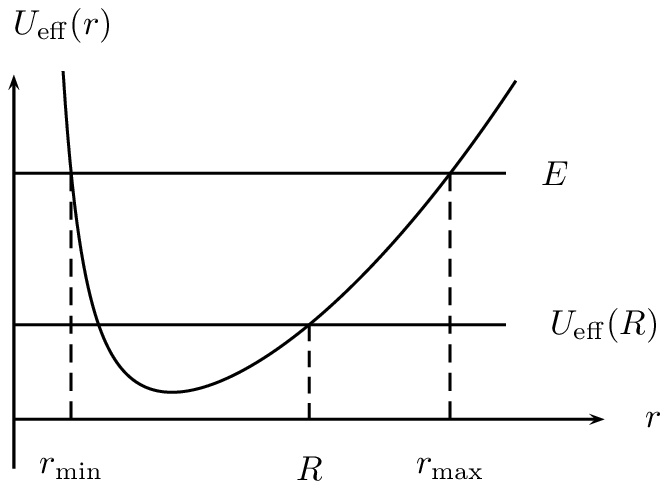}
\end{center}
\caption{Effective potential $U_{\rm eff}(r)$. An atom with
$E\geqslant U_{\rm eff}(R)$ is evaporated.} \label{fig:ueff}
\end{figure}

The confining potential $U_g$ being cylindrically symmetric, the
$z$ component $L$ of angular momentum is conserved. To study the
radial motion of an atom, one therefore uses the effective
one-dimensional potential
\begin{equation}
U_{\rm eff}(L,\,r)=U_g(r)+\frac{L^2}{2mr^2}
\end{equation}
which includes the centrifugal term. The radial coordinate $r(t)$
of the particle oscillates between the turning points $r_{\min}$
and $r_{\max}$. The atom is evaporated if and only if $r_{\min}
\leqslant R\leqslant r_{\max}$, or equivalently if its transverse
energy $E$ fulfills $E\geqslant U_{\rm eff}(R)$ (see Fig.
\ref{fig:ueff}). As a consequence, the evaporation criterion
depends on the two constants which define completely the
characteristics of the transverse trajectory: the total mechanical
transverse energy $E$ and the angular momentum $L$ along the
$z$-axis.

\subsubsection{Joint probability distribution of energy
and angular momentum}

We now determine, for a thermal distribution at temperature $T$ in
the potential $U_g$, the joint probability distribution $p\el$ for
an atom to have energy $E$ and angular momentum $L$. By definition
\begin{equation}
p\el=C\int  \,\mathrm{e}^{-\beta {\mathcal H}}\,\delta(E-{\mathcal
H})\,\delta(L-\ell_z)\,\dd^2 r\dd^2 p
\end{equation}
where  ${\mathcal H}(\rr,\pp)=p^2/(2m)+U_g(r)$,
$\ell_z=(\rr\times\pp)\cdot\ee_z$ and $C$ is a normalization
constant determined by $\int p\el\dd E\,\dd L =1$. The joint
probability can be readily recast in terms of the period of the
radial motion $T_r\el$ \cite{period}:
\begin{equation}
p\el=C'\ee^{-\beta E}T_r\el\,.
\end{equation}

\subsubsection{Harmonic confinement}

\begin{figure}
\begin{center}
\includegraphics[width=7.6cm]{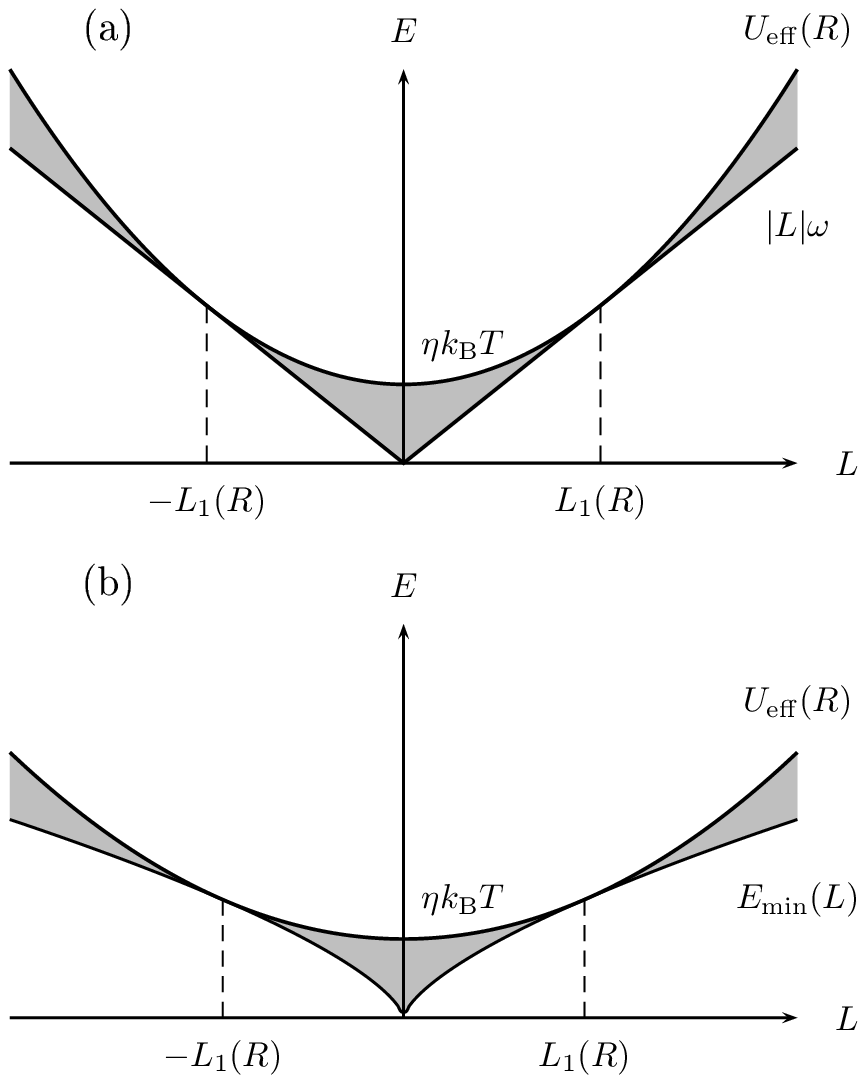}
\end{center}
\caption{Domain of integration $\mathcal{D}(R)$ in the plane
$\el$. $L_1(R)$ is the angular momentum of a particle having a
circular trajectory of radius $R$. (a): harmonic confinement (with
$L_1(R)=mR^2\omega$), (b): linear confinement (with
$L_1(R)=\sqrt{m\mu b R^3}$).} \label{fig:pel}
\end{figure}

For a harmonic confinement $U_g(r)=m\omega^2r^2/2$,
$T_r\el=\pi/\omega$ and $E\geqslant E_{\min}(L)\equiv|L|\omega$.
The latter condition just reflects that a finite angular momentum
requires a minimum energy which corresponds to the circular
motion. We readily obtain the exact expression for the joint
probability:
\begin{equation}
p_{\rm har}\el=\frac{\omega}{2(\kb T)^2}\ee^{-\beta E}
\Theta(E-E_{\min}(L)),
\end{equation}
where $\Theta$ is the Heaviside step function. The fraction of
remaining atoms is given by \cite{etah}:
\begin{equation}
\phi_{\rm har}(\eta)=\int_{\mathcal{D}(R)} p_H\el \dd E\dd L =
1-(\pi\eta)^{1/2}\mathrm{e}^{-\eta}\,. \label{eq:fh}
\end{equation}
The integration domain in Eq. (\ref{eq:fh}) is defined by
$\mathcal{D}(R)=\{E_{\min}(L)=|L|\omega\leqslant E\leqslant U_{\rm
eff}(R)\}$ and corresponds to the shaded area in the plane $\el$
depicted in Fig. \ref{fig:pel}. The subregion of $\mathcal{D}(R)$
such that $L\leqslant L_1(R)$ (resp. $L\geqslant L_1(R)$)
corresponds to trajectories not evaporated since $r_{\rm max}<R$
(resp. $r_{\rm min} >R$). The solid line in Fig. \ref{fig:Phi}
represents $\phi_{\rm har}(\eta)$. For small $R$ ($\eta \ll 1$),
few atoms have a sufficiently low angular momentum to be
evaporated and $\phi_{\rm har}\sim 1$. For large $R$ ($\eta \gg
1$) the evaporative loss is negligible because of the exponential
decay of the energy distribution and $\phi_{\rm har}\sim 1$.
Between those two limits, the fraction of remaining atoms has a
minimum for $\eta=\eta^*_{\rm har}=0.5$ with $\phi_{\rm
har}(\eta^*_{\rm har})\simeq 0.24$.

\subsubsection{Linear confinement}

\begin{figure}
\begin{center}
\includegraphics[width=7cm]{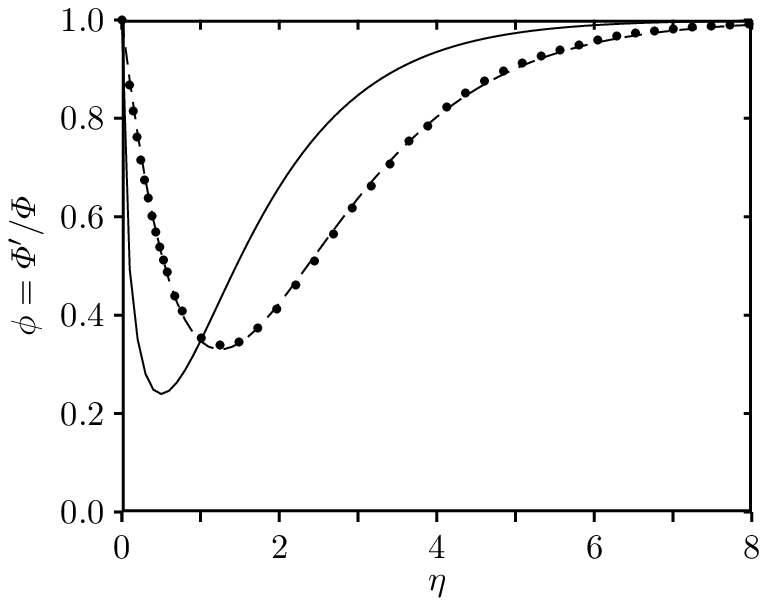}
\end{center}
\caption{Fraction $\phi(\eta)=\Phi'/\Phi$ of remaining atoms after
evaporation by one antenna, as a function of the evaporation
parameter $\eta$. Solid line: Harmonic confinement; dashed line:
Linear confinement, $\phi$ being calculated with the approximation
(\ref{eq:approx}); dots: Linear confinement, $\phi$ being
calculated by a Monte-Carlo simulation.} \label{fig:Phi}
\end{figure}

For linear confinement, the radial period $T_r\el$ has no simple
analytical expression. However, $T_r\el$ has a weak dependence on
the angular momentum $L$: for a given $E$, when $L$ varies from
$0$ (linear trajectory) to its maximal value (circular
trajectory), $T_r$ varies by less than $10\%$. We therefore make
the approximation $T_r\el\simeq T_r(E,0)\propto E^{1/2}$ and
approximate the joint distribution $p\el$ by:
\begin{equation}
p_{\rm lin}\el\simeq CE^{1/2}{\rm e}^{-\beta
E}\Theta(E-E_{\min}(L))\,. \label{eq:approx}
\end{equation}
The minimum value of $E$ for a given $L$ (energy of circular
motion) is now: $E_{\rm min}(L)=3(\mu b|L|)^{2/3}/(2m^{1/3})\,$.
The fraction of remaining atoms can be written as $\phi_{\rm
lin}(\eta)=\int_{\mathcal{D}(R)} p_L \el \dd E\dd L$ where the
integration domain is of the form $\mathcal{D}(R)=\{E_{\rm
min}(L)\leqslant E\leqslant U_{\rm eff}(R)\}$ (see Fig.
\ref{fig:pel}). The result of the numerical integration based on
the approximation (\ref{eq:approx}) is plotted as a dashed line in
Fig. \ref{fig:Phi}. The shape is qualitatively the same as that
for the harmonic confinement, but the minimum now occurs at
$\eta=\eta^*_{\rm lin}\simeq 1.25$ with $\phi_{\rm
lin}(\eta^*_{\rm lin})\simeq 0.33$.

To check the validity of the approximation (\ref{eq:approx}), we
performed a Monte-Carlo sampling of the atomic distribution in a
linear potential, followed by elimination of particles whose
trajectories crossed the circle of radius $R$. The result for
$\phi_{\rm lin}(\eta)$ is shown by dots in Fig. \ref{fig:Phi}. The
agreement is excellent.

It is convenient in practice to have a simple form for the
expression of $\phi_{\rm lin}(\eta)$. Following the functional
form of (\ref{eq:fh}), we fit the curve $\phi_{\rm lin}(\eta)$ by
a function of the form $1-a\eta^b{\rm e}^{-c\eta}$ where $a,b,c$
are adjustable parameters and find $a=1.65$, $b=1.13$ and
$c=0.92$. The curve obtained this way differs from $\phi_{\rm
lin}(\eta)$ by less than $2\%$. This simple expression is used in
\cite{prlrb2} to fit the experimental data, with the temperature
$T$ of the beam being the only adjustable parameter.

\subsection{Temperature change $T'/T$}

Consider $N$ atoms undergoing evaporation by one antenna. After
evaporation  the total transverse energy of the $N'=\phi(\eta)N$
remaining atoms is given by:
\begin{equation}
\eperp=N\int_{\mathcal{D}(R)}E\,p\el\,\dd E\, \dd L\equiv\xi N \kb
T g(\eta)\,,\label{eq:g}
\end{equation}
where $\xi=2$ for a harmonic potential and $\xi=3$ for a linear
potential. The dimensionless function $g(\eta)$ introduced in Eq.
(\ref{eq:g}) can be calculated analytically for the harmonic
confinement:
\begin{equation}
g_{\rm har}(\eta)=1-\frac{(\pi\eta)^{1/2}}{4}(3+2\eta){\rm
e}^{-\eta}\,.
\end{equation}
For a linear confinement, $g_{\rm lin}(\eta)$ is calculated
numerically thanks to the approximation (\ref{eq:approx}).

\begin{figure}
\begin{center}
\includegraphics[width=7cm]{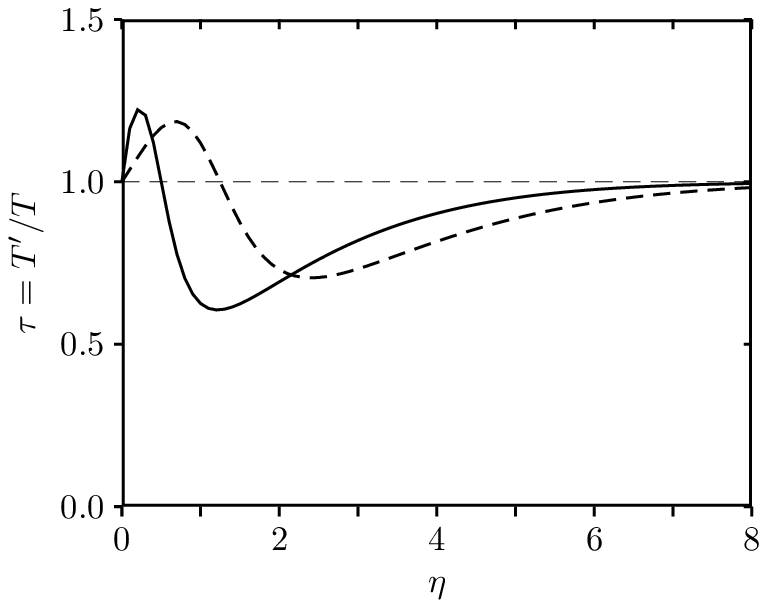}
\end{center}
\caption{Temperature change $\tau=T'/T$ after evaporation by one
antenna and rethermalization, as a function of $\eta$. Solid line:
harmonic confinement; dashed line: linear confinement.}
\label{fig:tpt}
\end{figure}

The total (transverse and longitudinal) energy of the remaining
atoms is $E'=\eperp+N'\kb T/2+N'm\bar{v}^2/2$. After
rethermalization (which occurs at constant total energy), one has
an equilibrium state with a temperature $T'$ and energy
$E'=(\xi+1/2)N'\kb T'+N'm\bar{v}^2/2$. The relative change of
temperature during this evaporation-rethermalization cycle is
therefore:
\begin{equation}
\tau(\eta)=\frac{T'}{T}=\frac{1}{2\xi+1}\bigg(1+2\xi\,\frac{g(\eta)}{\phi(\eta)}\bigg)\,.
\end{equation}
In Fig. \ref{fig:tpt}, $\tau(\eta)$ is plotted as a function of
$\eta$ (solid line: harmonic case, dashed line: linear case). For
low $\eta$ (approximately lower than the value $\eta^*$ for which
$\phi$ reaches a minimum), $T'>T$: since essentially low energy
atoms are evaporated, the remaining beam acquires a higher
temperature after thermalization. The opposite occurs for high
$\eta$: the beam is cooled by evaporation of high energy
particles.

\subsection{Variation of collision rate and phase-space density}

After an evaporation followed by a subsequent rethermalization,
the flux is multiplied by a factor $\phi$, and the beam
temperature is $T'$. We can then calculate, using Eqs.
(\ref{eq:psd}) and (\ref{eq:cr}), the new phase-space density
$\mathcal {D}'$ and the new collision rate $\coll'$.

\begin{figure}
\begin{center}
\includegraphics[width=7cm]{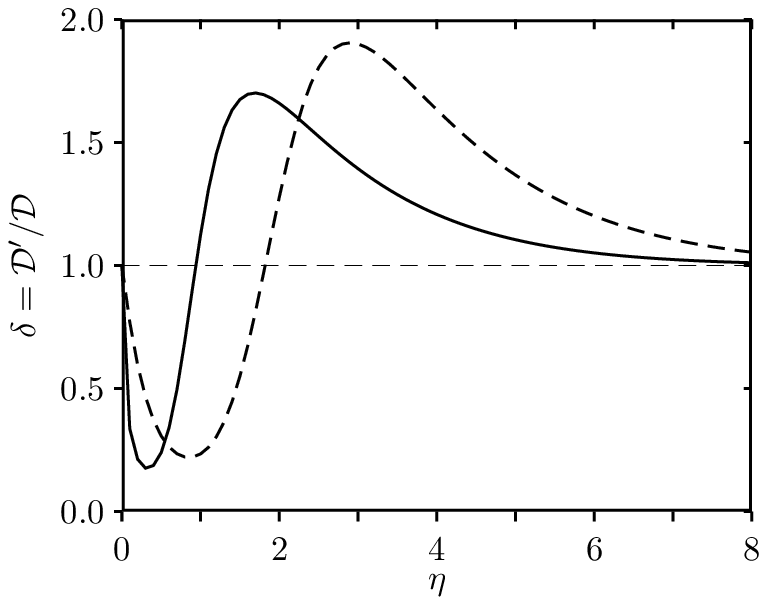}
\end{center}
\caption{Gain in phase-space density $\mathcal{D} '/\mathcal{D} $
after evaporation by one antenna and rethermalization, as a
function of $\eta$. Solid line: harmonic confinement; dashed line:
linear confinement.} \label{fig:RpR}
\end{figure}

The variation in phase-space density
$\delta(\eta)=\mathcal{D}'/\mathcal{D}$ is plotted as a function
of $\eta$ in Fig. \ref{fig:RpR} (solid line: harmonic case; dashed
line: linear case). For a linear confinement an increase of
$\mathcal{D} $ is obtained as soon as $\eta>2$, and for
$\eta\simeq 3.0$ a maximum gain in phase-space (by a factor 1.86)
can be achieved with a single antenna. The gain is less
significant in the case of a harmonic confinement, reflecting the
less favorable scaling law.

\begin{figure}
\begin{center}
\includegraphics[width=7cm]{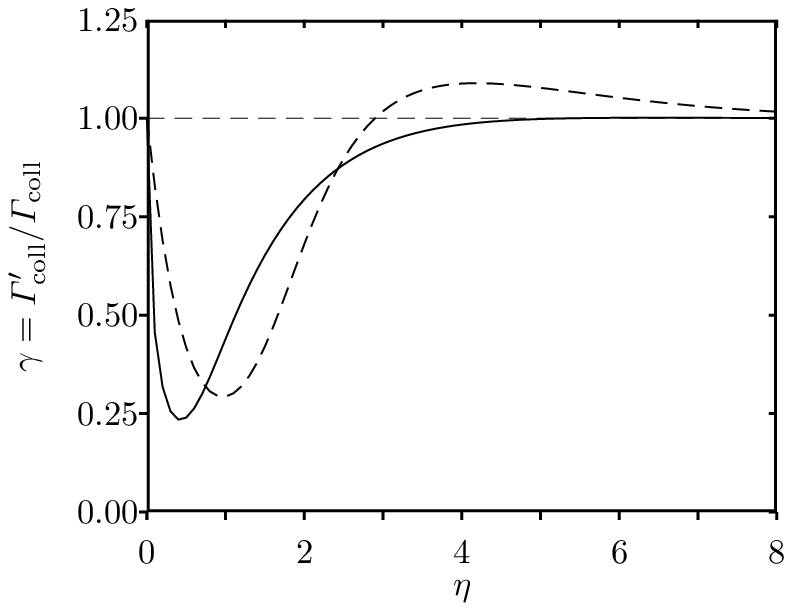}
\end{center}
\caption{Variation of the collision rate after evaporation by one
antenna and rethermalization, as a function of $\eta$. Solid line:
harmonic confinement; dashed line: linear confinement.}
\label{fig:gpg}
\end{figure}

In order for evaporative cooling to be efficient, the elastic
collision rate must not decrease during the process, otherwise the
time needed for rethermalization becomes prohibitively long. Fig.
\ref{fig:gpg} depicts the variation of collision rate
$\gamma(\eta)=\coll'/\coll$ for one evaporation cycle. The
striking result is that in the case of harmonic confinement the
collision rate cannot increase significantly ($\gamma$ reaches a
maximum value of $1.0015$ for $\eta=6.5$). This behaviour is due
to the two-dimensional geometry of the problem and does not hold
for a three-dimensional harmonic confinement. In contrast, for a
linear guide, and for $\eta$ sufficiently high, the collision rate
can increase by several percent at each cycle, creating the
possibility of \emph{runaway evaporation}. This salient feature of
the linear potential is well known for three-dimensional
evaporative cooling \cite{KetterleVanDruten,ketterleEvap}.

\subsection{Multi-radii evaporation}

The previous evaporation method has the disadvantage that for any
finite $\eta$, some atoms with large energy and angular momentum
are not evaporated. The contribution of these atoms to the energy
of the truncated distribution is not negligible, and therefore the
cooling efficiency of the evaporation is limited.

In order to improve this efficiency, one can evaporate all the
atoms whose trajectories lie, at least partially, outside of the
cylinder of radius $R$. To realize such an evaporation, one must
use a ``continuum" of evaporation radii $R_e \in [ R , \infty ]$.
In practice, this can be achieved by scanning the radio-frequency
$\nu_{\rm RF}$ from $U_g(R)/h$ to a value large compared to
$k_{\rm B} T/h$, above which the population of atoms is
exponentially small. The scanning rate must be large compared to
the inverse of the time $t_{\rm ev}$ spent by an atom in the range
of efficiency of an antenna, and small compared to the inverse of
the mean radial period $T_r\el$. For the typical parameters of the
experiment described in \cite{prlrb2}, a scanning frequency of
50~Hz would fulfill both criteria.

\begin{figure}
\begin{center}
\includegraphics[width=7cm]{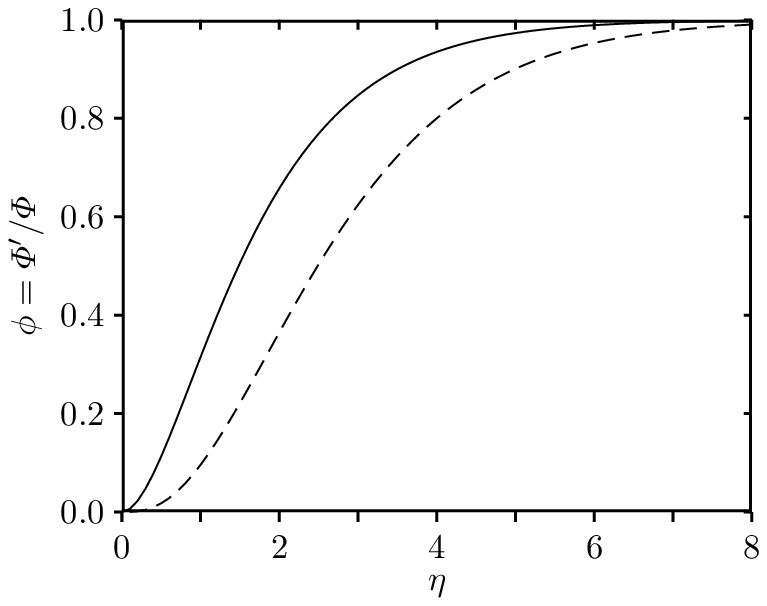}
\end{center}
\caption{Fraction $\phi(\eta)=\Phi'/\Phi$ of remaining atoms after
evaporation by one antenna, as a function of $\eta$, for the
multi-radii evaporation scheme. Solid line: harmonic confinement;
dashed line: linear confinement.} \label{fig:phimultiradii}
\end{figure}

\begin{figure}
\begin{center}
\includegraphics[width=7cm]{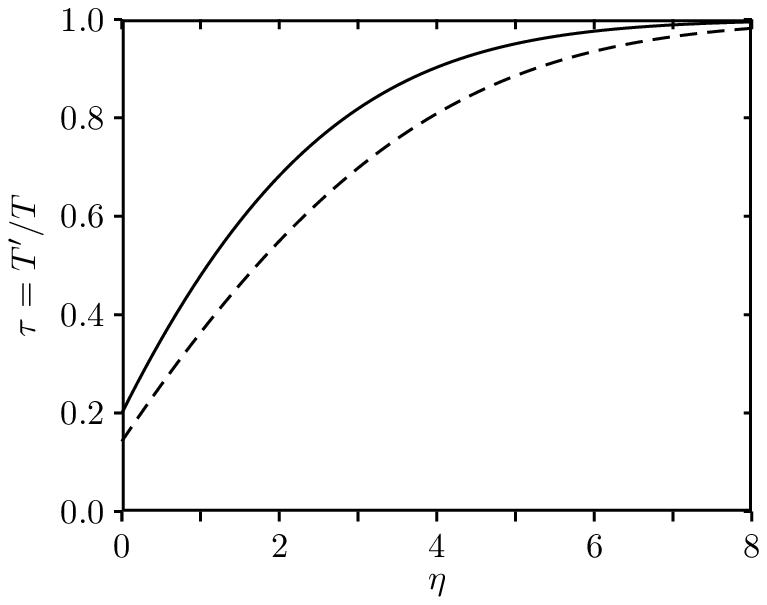}
\end{center}
\caption{Temperature change $\tau=T'/T$ after evaporation by one
antenna and rethermalization, as a function of $\eta$, for the
multi-radii evaporation scheme. Solid line: harmonic confinement;
dashed line: linear confinement.} \label{fig:Tmultiradii}
\end{figure}

\begin{figure}
\begin{center}
\includegraphics[width=7cm]{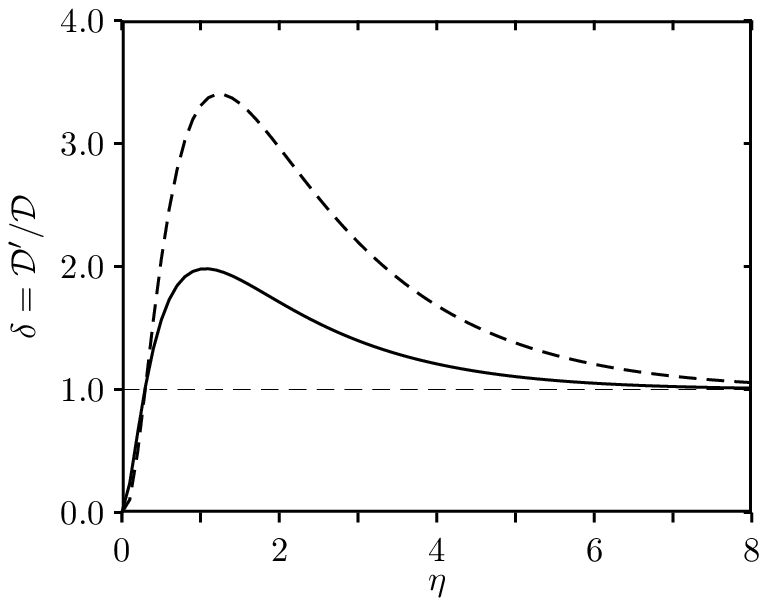}
\end{center}
\caption{Gain in phase space density after evaporation by one
antenna and rethermalization, as a function of $\eta$, for the
multi-radii evaporation scheme. Solid line:harmonic confinement;
dashed line: linear confinement.} \label{fig:Dmultiradii}
\end{figure}

\begin{figure}
\begin{center}
\includegraphics[width=7cm]{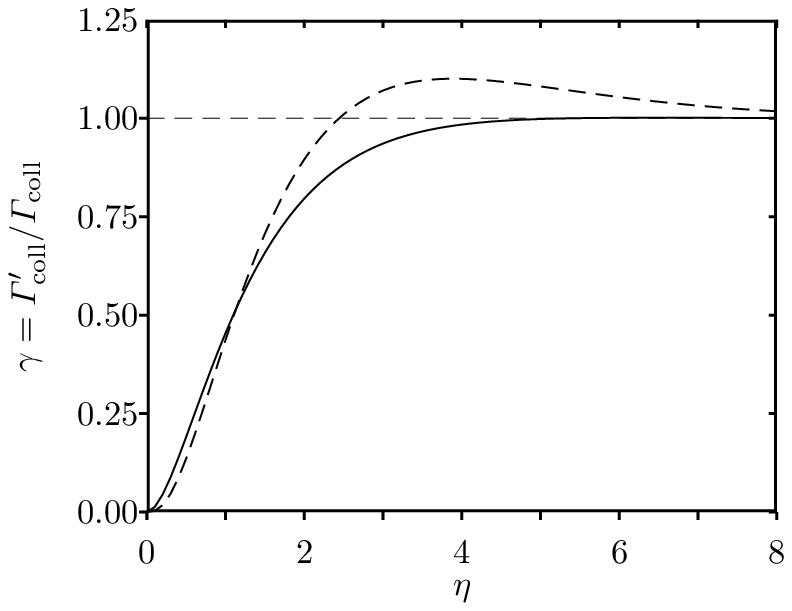}
\end{center}
\caption{Variation of the collision rate after evaporation by one
antenna and rethermalization, as a function of $\eta$, for the
multi-radii evaporation scheme. Solid line:harmonic confinement;
dashed line: linear confinement.} \label{fig:Gmultiradii}
\end{figure}

One can calculate, in  a way similar to the calculations of the
previous subsections, the change in atomic flux
$\phi(\eta)=\Phi'/\Phi$, in temperature $\tau(\eta)=T'/T$, in
phase-space density $\delta(\eta)=\mathcal{D} '/\mathcal{D}$, and
in collision rate $\gamma(\eta)=\coll'/\coll$, for this improved
evaporation scheme. The domain of integration is that shown in
Fig. \ref{fig:pel} but reduced to the angular momentum lower than
$L_1(R)$. In the case of harmonic confinement, the calculations
can be done explicitly and yield the following results:
\begin{eqnarray}
\phi(\eta)&=&1-\ee^{-2\eta}-(\pi\eta)^{1/2}\ee^{-\eta}{\rm
erf}(\eta^{1/2}),\nonumber \\
 g(\eta)=&1&-\,(1+\eta/2)\ee^{-2\eta}\nonumber
\\&&-\,(3/4+\eta/2)(\pi\eta)^{1/2}{\rm e}^{-\eta}{\rm
erf}(\eta^{1/2})\,
\end{eqnarray}
where ${\rm erf}(x)=2\pi^{-1/2}\int_0^x\,\exp(-t^2)\,\dd t$ is the
error function. The complete results for $\phi$, $\tau$, $\gamma$
and $\delta$ and for both types of confinement are plotted in
Figs. \ref{fig:phimultiradii} to \ref{fig:Gmultiradii}. Obviously,
with this new evaporation scheme, the number of atoms remaining
vanishes when $\eta\rightarrow 0$. The temperature change (Fig.
\ref{fig:Tmultiradii}) is now a monotonous function of $\eta$ and
for $\eta\rightarrow 0$, in the harmonic (resp. linear) case one
has $\tau\rightarrow 1/5$ (resp. $1/7$) since only the
longitudinal kinetic energy contributes to the energy of the
remaining particles.

For a linear potential, the gain in phase-space density without
decrease in collision rate is 2.56 (at $\eta=2.5$) as compared to
1.86 (at $\eta=3.0$) in the single radius evaporation scheme.

\section{Evaporation with two antennas}
\label{sec:2ant}

The presence of elastic collisions in the atomic beam can be
probed with a two-antenna experiment \cite{prlrb2}. A first
antenna with frequency $\nu_1$ evaporates the atoms with an
evaporation parameter $\eta_1=h\nu_1/(\kb T)$, where $T$ is the
initial temperature of the atoms. The remaining flux is
$\Phi'=\phi(\eta_1)\Phi$. A second antenna with frequency
$\nu_2=\eta_2\kb T/h$, placed at a distance $d$ downstream, is
used to probe the distribution of atoms after this first
evaporation stage. The flux after this second antenna is denoted
$\Phi''$. In the absence of collisions between the two antennas,
the distribution remains out of equilibrium. In particular, if
$\nu_2=\nu_1$, the second antenna cannot remove extra atoms, since
all atoms whose energy and angular momentum corresponded to the
evaporation criterion have already been removed by the first
antenna. But even a partial rethermalization leads to some extra
losses from the second antenna, since elastic collisions provide a
redistribution of energy and angular momentum. If the beam
completely rethermalizes during its propagation between the two
antennas, we find a spectrum similar to that of Fig. \ref{fig:Phi}
revealing the new temperature $T'$.

\subsection{In absence of collisions}

\begin{figure}
\begin{center}
\includegraphics[width=7cm]{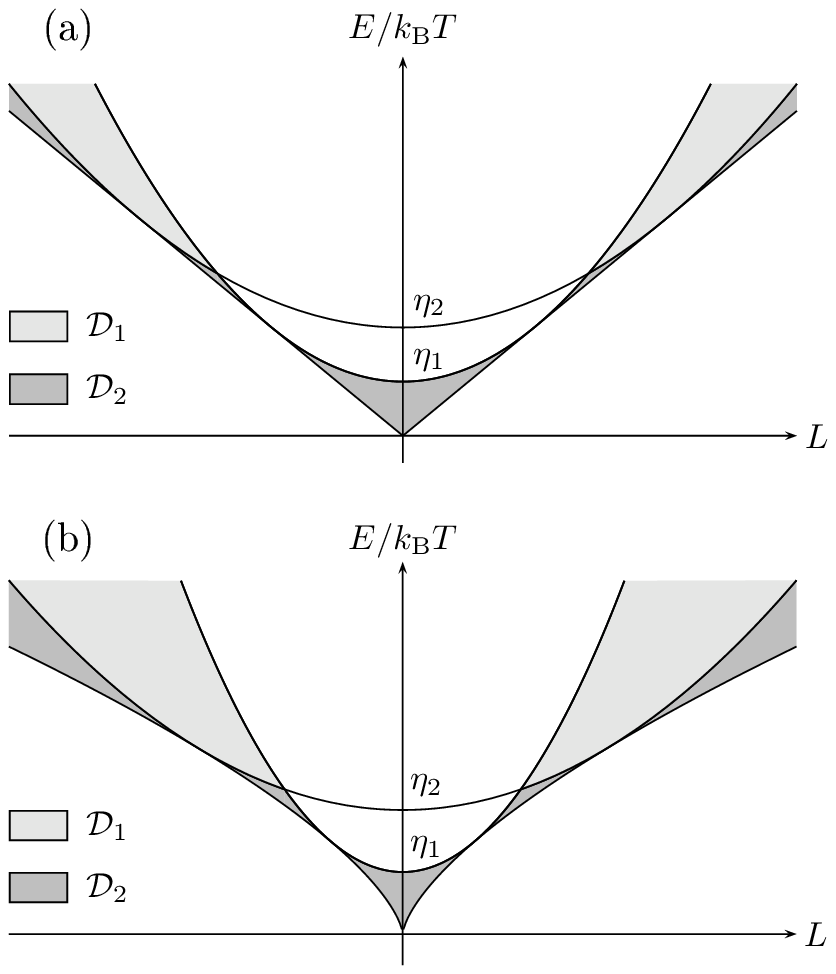}
\end{center}
\caption{The integration domains in the plane $\el$ involved in
the calculation of $\Phi''/\Phi'$ in the collisionless regime.
(a): Harmonic confinement; (b): Linear confinement.}
\label{fig:pelcl}
\end{figure}

\begin{figure}
\begin{center}
\includegraphics[width=7cm]{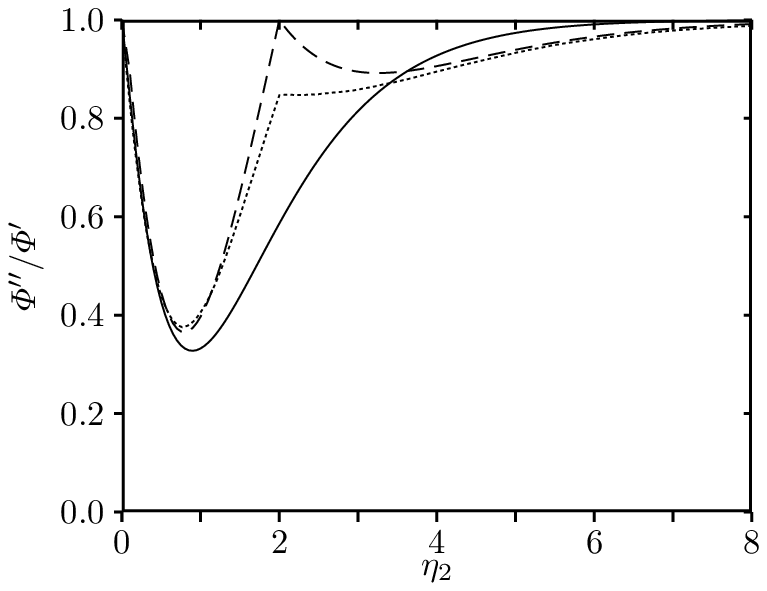}
\end{center}
\caption{Fraction $\Phi''/\Phi'$ of the flux remaining after the
second antenna as a function of its parameter $\eta_2$. The first
antenna is driven at a frequency $\nu_1=\eta_1 k_BT/h$ with
$\eta_1=2=1.6\eta^*$. The dashed line corresponds to the
collisionless regime; the dotted line to a mean number of
collisions between antennas $N_c=1.1$; the solid line to full
rethermalization. These curves are calculated with a numerical
simulation, in the case of a linear confinement.}
\label{fig:phicl}
\end{figure}

We calculate here the fraction $\Phi''/\Phi'$ in the limit where
rethermalization does not occur between the two antennas, {\it
i.e.} when the inequality $\coll .d \ll \bar{v}$ is fulfilled.
After evaporation by the first antenna at $\eta_1$ (corresponding
to an evaporation radius $R_1$), the out of equilibrium joint
probability distribution $p_1\el$ of $\el$ is given by
\begin{equation}
p_1\el= p\el \times  \Theta(U_{\rm eff}(R_1)-E)\,. \label{eq:ooe}
\end{equation}
It is non-zero only in the domain $\mathcal{D}_1\cup
\mathcal{D}_2$ of Fig. \ref{fig:pelcl}. The fraction of remaining
atoms $\Phi''/\Phi'$ is therefore given by integrating
(\ref{eq:ooe}) over the domain $\mathcal{D}_2=\{
E_{\min}(L)\leqslant E \leqslant \min(U_{\rm eff}(R_1),U_{\rm
eff}(R_2)) \}$:
\begin{equation}
\frac{\Phi''}{\Phi'}=\frac{\displaystyle \int_{\mathcal{D}_2} p_1
\el\, \dd E\, \dd L}{\displaystyle
\int_{\mathcal{D}_1\cup\mathcal{D}_2}  \!\!\!\!\!\!\!\!\! p_1
\el\, \dd E\, \dd L}\,.
\end{equation}
The result of the numerical integration is shown in Fig.
\ref{fig:phicl} as a dashed line, for the specific case of a
linear confinement with $\eta_1=2$. As expected, the fraction
$\Phi''/\Phi'$ reaches $1$ for $\eta_2=2$.

\subsection{Rethermalization}

If rethermalization occurs over a distance $d$, that is if the
collision rate is such that: $\coll .d \gg \bar{v}$, the second
antenna evaporates a beam of temperature $T'=\tau(\eta_1)T$ at
thermal equilibrium. The fraction $\Phi''/\Phi'$ is therefore:
\begin{equation}
\frac{\Phi''}{\Phi'}=\phi\left ( \frac{h\nu_2}{\kb T'} \right  ) =
\phi\left ( \frac{\eta_2}{\tau(\eta_1)} \right  )\,.
\end{equation}
This curve is plotted as a solid line on Fig. \ref{fig:phicl}. The
large difference between the collisionless and collisional regimes
makes it easy to detect experimentally, by scanning $\nu_2$ in the
vicinity of $\nu_1$, the presence or the absence of collisions
within the beam \cite{prlrb2}.

\subsection{Number of collisions required to thermalize}
\label{sec:numcol}

\begin{figure}
\begin{center}
\includegraphics[width=7cm]{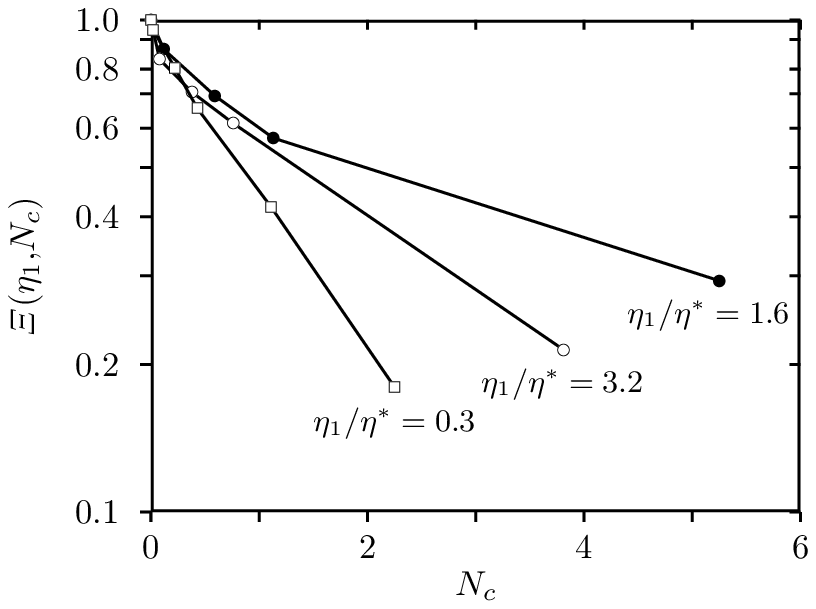}
\end{center}
\caption{Distance from equilibrium $\Xi(\eta_1,N_c)$ (see text) as
a function of the mean number of collisions $N_c$ between the two
antennas, for different values of $\eta_1$. These data have been
obtained by a Monte-Carlo simulation performed for a linear
confinement. The relaxation towards equilibrium is clearly
exponential for $\eta_1/\eta^*=0.3$, and approximately exponential
for $\eta_1/\eta^*=3.2$. For $\eta_1/\eta^*=1.6$, the first
antenna puts the gas in a strongly out-of-equilibrium state, and
the relaxation is not exponential.} \label{fig:decay}
\end{figure}

In order to give a quantitative theoretical account for the
distance $d$ needed to rethermalize in a two-antenna experiment,
we have developed a numerical simulation of the motion of the
atoms in the magnetic guide for both linear and harmonic
confinements. We use a molecular dynamics simulation \cite{Bird}
taking into account the evaporation criterion.

The simulation is performed in dimensionless units. The unit
length $r_u$ is defined by $U_g(r_u)-U_g(0)=(\langle
U_g\rangle_T-U_g(0))/2$, the distance between the two antennas is
$d=5000\;r_u$. The mean velocity is 10 times the thermal velocity
$(k_BT/m)^{1/2}$. The phase-space variables of each particle are
evolved by advancing the position and the velocity at discrete
time steps $\Delta t$ according to a second-order symplectic
integration \cite{jackson}. Using the cylindrical symmetry we
restrict the evolution to the first quarter ($x>0$, $y>0$) of
space with reflecting walls at planes $x=0$ and $y=0$. Binary
elastic collisions are taken into account using a boxing technique
\cite{boxing}. The constant cross section used to calculate the
probability of collisions is adjusted to the desired number of
collisions between the two antennas. Simulations have been
performed with $1.5\times 10^6$ particles and $5.4\times 10^4$
boxes. The time step $\Delta t$ is chosen to be small with respect
to the typical collision timescale and the typical period of
oscillation.

We simulate a two-antenna experiment in which the first antenna is
operated at a fixed $\eta_1$. The mean number of collisions per
atom during the propagation between the two antennas is denoted
$N_c$. The second antenna is operated at $\eta_2$, and we
calculate $\Phi''/\Phi'$ as a function of $\eta_2$. An example of
such a curve, for a linear confinement, and with $N_c=1.1$, is
plotted as a dotted line in Fig. \ref{fig:phicl}.

We then infer the ratio
$f(\eta_1,N_c)=\Phi''/\Phi'(\eta_2=\eta_1)$ as a function of
$N_c$. For a collisionless propagation ($N_c=0$, dashed line in
Fig. \ref{fig:phicl}), we have $f(\eta_1,0)=1$. For a complete
rethermalization, ($N_c\gg 1$, solid line in Fig.
\ref{fig:phicl}), the ratio tends to the limit
$f_\infty(\eta_1)=\phi[\eta_1/\tau(\eta_1)]$.

For a linear confinement, we depict in Fig. \ref{fig:decay} the
normalized quantity
$$\Xi(\eta_1,N_c)=\frac{f(\eta_1,N_c)-f_\infty(\eta_1)}{1-f_\infty(\eta_1)}\,,$$
which measures the distance from equilibrium, as a function of the
mean number of collisions $N_c$. For a given $\eta_1$ and in the
absence of collisions $N_c=0$, $\Xi=1$. The relaxation of $\Xi$
towards zero with increasing number of collisions $N_c$ reflects
the rethermalization process. For $\eta_1<0.3\eta^*$ or
$\eta_1>2.5\eta^*$, $\Xi$ decays approximately in an exponential
manner with $N_c$: $\Xi\simeq \exp[-N_c/n_c(\eta_1)]$. As an
example we find $n_c(4)=2.5\pm 0.5$ for the linear confinement and
$n_c(2)=1.3\pm 0.2$ for the harmonic confinement. For intermediate
values of the ratio $\eta_1/\eta^*$, we obtain a non-exponential
decay (see Fig. \ref{fig:decay} for $\eta_1/\eta^*=1.6$). In this
range $0.3<\eta_1/\eta^*<2.5$, more than $50\%$ of the atoms are
evaporated by the first antenna, leading to a state of gas that is
very far from equilibrium.

This calibration by numerical simulation allows one to determine
the collision rate by following the thermalization process for
different distances between the two antennas \cite{prlrb2}.

\section{Evaporation with many antennas}
\label{sec:Mant}

The realization of a magnetically guided atomic beam in the
collisional regime has recently been performed \cite{prlrb2}. The
phase-space density of this beam is on the order of $2\times
10^{-8}$. To gain the 8 orders of magnitude required to reach
degeneracy one can implement evaporative cooling by means of
several successive antennas. In this section we assume that the
frequency of each antenna is such that the parameter $\eta$ is
constant throughout the evaporation.

The number of antennas ${\cal N}(\eta)$ required to achieve a gain
$\Delta$ in phase-space is, for a given value of $\eta$, ${\cal
N}(\eta)=\ln(\Delta)/\ln(\delta(\eta))$. We want to minimize the
guide length needed to reach degeneracy. This length $\ell(\eta)$
can be evaluated from the initial collision rate of the beam
$\coll^0$. Between the antennas $n$ and $n+1$, the collision rate
is $\coll^0\gamma^n$. If the number of collisions required for
full thermalization between two successive antennas is $N_0$
($N_0\gg n_c(\eta)$), the distance between antennas $n$ and $n+1$
has to be larger than $\bar{v}N_0/(\coll^0\gamma^n)$. Therefore
the total length is

\begin{eqnarray}
\ell(\eta)&\geqslant&\frac{\bar{v}
N_0}{\coll^0}\bigg(1+\gamma^{-1}+ ...
+\gamma^{-{\cal N}+1} \bigg)\nonumber \\
 &\geqslant&\frac{\bar{v}
N_0}{\coll^0}\frac{1-\gamma^{-{\cal N}}}{1-\gamma^{-1}}.
\label{eq:ell}
\end{eqnarray}

For a given value of $\Delta$, there exists an optimal choice
$\tilde{\eta}$ for the parameter $\eta$ such that $\ell$ is
minimized. For this optimization we assume that $N_0$ does not
depend on $\eta$. The results are shown, for both types of
confinement and for the two evaporation schemes, in Table
\ref{tab:evap}. The evaporation is much less efficient for the
harmonic confinement, which reflects the absence of runaway
evaporation for this type of trap. For the linear confinement, one
sees that the multi-radii evaporation scheme leads to a slightly
shorter evaporation length, at the expense of a reduced output
flux. However a more careful study shows that for comparable
output fluxes, the two schemes require approximately the same
length and antenna numbers. In the case of harmonic confinement,
since the collision rate cannot increase significantly, minimizing
the evaporation length requires that one operates at a large
$\eta$ (where $\gamma\simeq 1$). For such high $\eta$, the
difference between the two evaporation schemes is negligible, as
can be seen in Table~\ref{tab:evap}.

\begin{table}[t]
\begin{center}
\small
\begin{tabular}{|c||c|c|c|c|}
\hline \rule[-0.3cm]{0mm}{0.8cm} Scheme &
\multicolumn{2}{|c|}{Single radius} & \multicolumn{2}{|c|}{Multi-radii} \\
\hline  \rule[-0.3cm]{0mm}{0.8cm} Trap & Har & Lin & Har &
Lin\\
\hline \hline \rule[-0.3cm]{0mm}{0.8cm}
$\ell_{\min}$ & $211$ & $12.0$ & $211$ & $10.5$ \\
\hline \rule[-0.3cm]{0mm}{0.8cm}
$\tilde{\eta}$ & $4.6$ & $4.3$ & $4.6$ & $3.6$ \\
\hline \rule[-0.3cm]{0mm}{0.8cm}
$\mathcal{N}$ & $142$ & $42$ & $142$ & $29$ \\
\hline \rule[-0.3cm]{0mm}{0.8cm}
$\Phi_f/\Phi_i$ & $3.9\,.\, 10^{-3}$ & $5.4\,.\, 10^{-4}$ & $3.9\,.\, 10^{-3}$ & $1.3\,.\,10^{-4}$ \\
\hline \rule[-0.3cm]{0mm}{0.8cm}
$T_f/T_i$ & $6.9\,.\, 10^{-5}$ & $6.0\,.\, 10^{-4}$ & $6.9\,.\, 10^{-5}$ & $3.8\,.\, 10^{-4}$ \\
\hline
\end{tabular}
\normalsize
\end{center}
\caption{Parameters of the optimized (minimum length) evaporation
ramp for harmonic (Har) and linear (Lin) confinements, for the two
schemes of evaporation, with a total gain in phase-space density
of $\Delta=10^8$: $\ell_{\min}$, evaporation length (in units of
$\bar{v}N_0/\coll^0$); $\tilde{\eta}$, parameter $\eta$ for
minimal evaporation length; $\mathcal{N}$, number of antennas
needed; $\Phi_f/\Phi_i$, total flux reduction due to evaporation;
$T_f/T_i$, total temperature reduction due to evaporation. }
\label{tab:evap}
\end{table}

In this paper, two competing effects that should be considered in
a more realistic model of the evaporation ramp have been
neglected. First, the real shape of the semi-linear potential
(\ref{potential}) should be taken into account. Indeed, even if at
the beginning of the evaporation ramp the temperature is high
enough for the linear potential to be a good approximation
($\alpha\ll 1$), as the temperature is reduced the effect of the
bias field cannot be neglected anymore when $\alpha\gtrsim 1$.
This effect may lead to a longer evaporation length since, for the
harmonic confinement, the evaporative cooling is less efficient.
The second effect, neglected by the classical description used in
this paper, is the bosonic stimulation close to degeneracy, which
might enhance the efficiency of the last evaporation steps.

\section{Conclusion}
\label{sec:conclusion}

We have studied theoretically the discrete-step evaporative
cooling of a magnetically guided atomic beam. First, the action of
a single antenna on the beam is analyzed in detail. Then we
clarify how two-antenna experiments can lead (i) to the
characterization of the collisional regime of the beam and (ii) to
the quantitative estimate of the collision rate, by studying the
number of collisions needed for thermalization of the truncated
distribution function. Finally, the problem of optimizing the
evaporation ramp with many antennas is addressed. We find that
gains in phase-space density of $10^8$ require a total of at least
$\sim 30$ evaporation steps.

\begin{acknowledgement}
We are indebted to J. Dalibard for useful comments and to P.~Cren
for the development of a preliminary version of the numerical
simulation. We acknowledge stimulating discussions with J.~M.
Vogels, K.~J. G\"unter, K. Kozlowski and the ENS laser cooling
group. We thank D.~Roberts for careful reading of the manuscript.
This work was partially supported by the CNRS, the R\'egion
Ile-de-France, the French Ministry of Research, the Coll\`{e}ge de
France, the Ecole Normale Sup\'erieure, the Bureau National de la
M\'etrologie, the University of Paris VI and the D\'el\'egation
G\'en\'erale de l'Armement.
\end{acknowledgement}


\begin{thebibliography}{99}
\bibitem{quadnot}
If one superimposes a constant magnetic field over a
three-dimensional quadrupole, there is always a point where the
modulus of the magnetic field vanishes. By contrast, if the
constant magnetic field has a non zero component on the axis of
the two-dimensional quadrupole configuration of a magnetic guide,
the resulting field never vanishes.
\bibitem{period}
$T_r\el=2\int_{r_{\rm min}}^{r_{\rm max}} \dd r/\dot{r}$
 where $\dot r =
\sqrt{2/m}[E-U_g(r)-L^2/(2mr^2)]^{1/2}$ is the radial velocity of
the atom.
\end{thebibliography}
\end{document}